\pdfoutput=1
\documentclass{article}

\usepackage{arXiv}
\usepackage{graphicx}
\usepackage{amsfonts,amssymb}
\usepackage{enumerate}
\usepackage{amsmath,bm}
\usepackage{times}

\title{Question Answering over Knowledge Base with Neural Attention Combining Global Knowledge Information}
\author{
Yuanzhe Zhang\footnotemark[1], Kang Liu\footnotemark[1], Shizhu He\footnotemark[1], Guoliang Ji\footnotemark[1], Zhanyi Liu\footnotemark[2], Hua Wu\footnotemark[2], Jun Zhao\footnotemark[1]\\
\footnotemark[1] Institute of Automation, Chinese Academy of Sciences\\
\{yzzhang, kliu, shizhu.he, guoliang.ji, jzhao\}@nlpr.ia.ac.cn\\
\footnotemark[2] Baidu Inc.\\
\{liuzhanyi, wu\_hua\}@baidu.com
}

\begin{document}

\maketitle

\begin{abstract}
  With the rapid growth of knowledge bases (KBs) on the web, how to take full advantage of them becomes increasingly important. Knowledge base-based question answering (KB-QA) is one of the most promising approaches to access the substantial knowledge. Meantime, as the neural network-based (NN-based) methods develop, NN-based KB-QA has already achieved impressive results. However, previous work did not put emphasis on question representation, and the question is converted into a fixed vector regardless of its candidate answers. This simple representation strategy is unable to express the proper information of the question. Hence, we present a neural attention-based model to represent the questions dynamically according to the different focuses of various candidate answer aspects. In addition, we leverage the global knowledge inside the underlying KB, aiming at integrating the rich KB information into the representation of the answers. And it also alleviates the out of vocabulary (OOV) problem, which helps the attention model to represent the question more precisely. The experimental results on WEBQUESTIONS demonstrate the effectiveness of the proposed approach.
\end{abstract}

\section{Introduction}

As the amount of the knowledge bases (KBs) grows, people are paying more attention to seeking effective methods for accessing these precious intellectual resources. There are several tailor-made languages designed for querying KBs, such as SPARQL \cite{prud2008sparql}. However, to handle such query languages, users are required to not only be familiar with the particular language grammars, but also be aware of the vocabularies of the KBs. By contrast, knowledge base-based question answering (KB-QA) \cite{unger2014introduction}, which takes natural language as query language, is a more user-friendly solution, and has become a research focus in recent years.

The goal of KB-QA is to automatically return answers from the KB given natural language questions. There are two mainstream research directions for this task, i.e., semantic parsing-based (SP-based) \cite{zettlemoyer2005learning,zettlemoyer2009learning,kwiatkowski2013scaling,cai2013large,berant2013semantic,yih2015semantic} and information retrieve-based (IR-based) \cite{yao2014information,bordes2014open,bordes2014question,dong2015question,bordes2015large} methods. SP-based methods usually focus on constructing a semantic parser that could convert natural language questions into structured expressions like logical forms. IR-based methods are more like to search answers from the KB based on the information conveyed in the questions. Here, ranking techniques are often adopted to make correct selections from candidate answers. In general, IR-based methods are easier and more flexible to implement. \cite{dong2015question,bordes2015large} have proven that IR-based methods could acquire competitive performance compared with SP-based methods through the experiments conducted over Freebase \cite{bollacker2008freebase}.

Recently, with the progress of deep learning, neural network-based (NN-based) methods have been introduced to the KB-QA task \cite{bordes2014open}. They belong to IR-based methods. Different from previous methods, NN-based methods represent both the questions and the answers as semantic vectors. Then the complex process of KB-QA could be converted into a similarity matching process between an input question and its candidate answers in a semantic space. The candidates with the highest similarity score will be considered as the final answers. Because they are adaptive and robust, NN-based methods have attracted more and more attention, and this paper also focus on using neural networks to answer questions over knowledge base.

In NN-based methods, the crucial step is to compute the similarity score between a question and a candidate answer, where the key is to learn their representations.
Previous methods put more emphasis on learning representations of the answer end. For example, \cite{bordes2014question} considers the importance of the subgraph of the candidate answers. \cite{dong2015question} makes use of the context and the type of the answers. By contrast, the representation methods of the question end are oligotrophic.
Existing approaches often represent a question into a single vector using a simple bag-of-words (BOW) model \cite{bordes2014open,bordes2014question}, whereas its relatedness to the answer end is neglected. We argue that a question should be represented differently according to the different focuses of various answer aspects\footnote{An answer aspect could be the answer entity itself, the answer type, the answer context, etc.}.

Take question ``Who is the president of France?'' and one of its candidate answers ``Francois Hollande'' as an example. When dealing with the answer entity {\tt Francois Holland}, ``president'' and ``France'' in the question is more focused, and the question representation should bias towards the two words. While facing the answer type {\tt /business/board\_member}, ``Who'' should be the most prominent word. Obviously, this is an attention mechanism, which reflects how the focus of answer aspects could influence the representation of the question.

When learning the representations of the questions, we should make proper use of each word in the question according to different attention of each aspect of the candidate answer, instead of simply compressing them into a fixed vector. We believe that such kind of representations are more expressive. \cite{dong2015question} represents questions using three CNNs with different parameters when dealing with different answer aspects including the answer path, the answer context and the answer type. We think simply selecting three independent CNNs is mechanical and inflexible. Thus, we go one step further, and propose an attention-based neural network to perform question answering over KB. Different to \cite{dong2015question}, we represent the question differently according to different answer resources, not allowing them sharing the same network as \cite{dong2015question} does. For instance, {\tt /business/board\_member} and {\tt /location/country} are both answer types, but the question representation will be different according to their different attention in our method.

On the other hand, we notice that the representations of the KB resources (entities and relations) are also limited in previous work. To be specific, they are often learned barely on the QA training data, which results in two limitations. 1) \textit{The deficiency of the global information of the KB}. The previous methods merely utilize the answer-related part in the KB, i.e., answer path and answer context \cite{bordes2014question,dong2015question}, to learn the representations of KB resources. The global information of the KB is completely ignored. For example, if question-answer pair $(q,a)$ appears in the training data, and the global KB information implies us that $a'$ is similar to $a$ \footnote{The complete KB is able to offer this kind of information, e.g., $a$ and $a'$ share massive context.}, denoted by ($a \sim a'$), then $(q,a')$ is more probable to be right. However, current QA training mechanism cannot guarantee ($a \sim a'$) could be learned. 2) \textit{The problem of out of vocabulary (OOV)}. Due to the limited coverage of the training data, the OOV problem is common while testing, and many answer entities in testing candidate set have never been seen before. In this scenario, the representation of such unseen KB resources could not be learned precisely. The attention of these resources become the same because they shared the same OOV embedding, and this will do harm to the proposed attention model. To tackle these two problems, we additionally incorporates KB itself as training data for training embeddings besides original question-answer pairs. In this way, the global structure of the whole knowledge could be captured, and the OOV problem could be alleviated naturally.

In summary, the contributions of this paper are as follows.
\begin{enumerate}[1)]
\item We present a novel attention-based NN model tailored to the KB-QA task, which considers the influence of the answer aspects for representing questions.
\item We leverage the global KB information, aiming at representing the answers more precisely. It also alleviates the OOV problem.
\item The experimental results on the open dataset WEBQUESTIONS demonstrate the effectiveness of the proposed approach.
\end{enumerate}

\section{Overview}
\begin{figure}[!ht]
	\centering
	\includegraphics[width=0.99\linewidth]{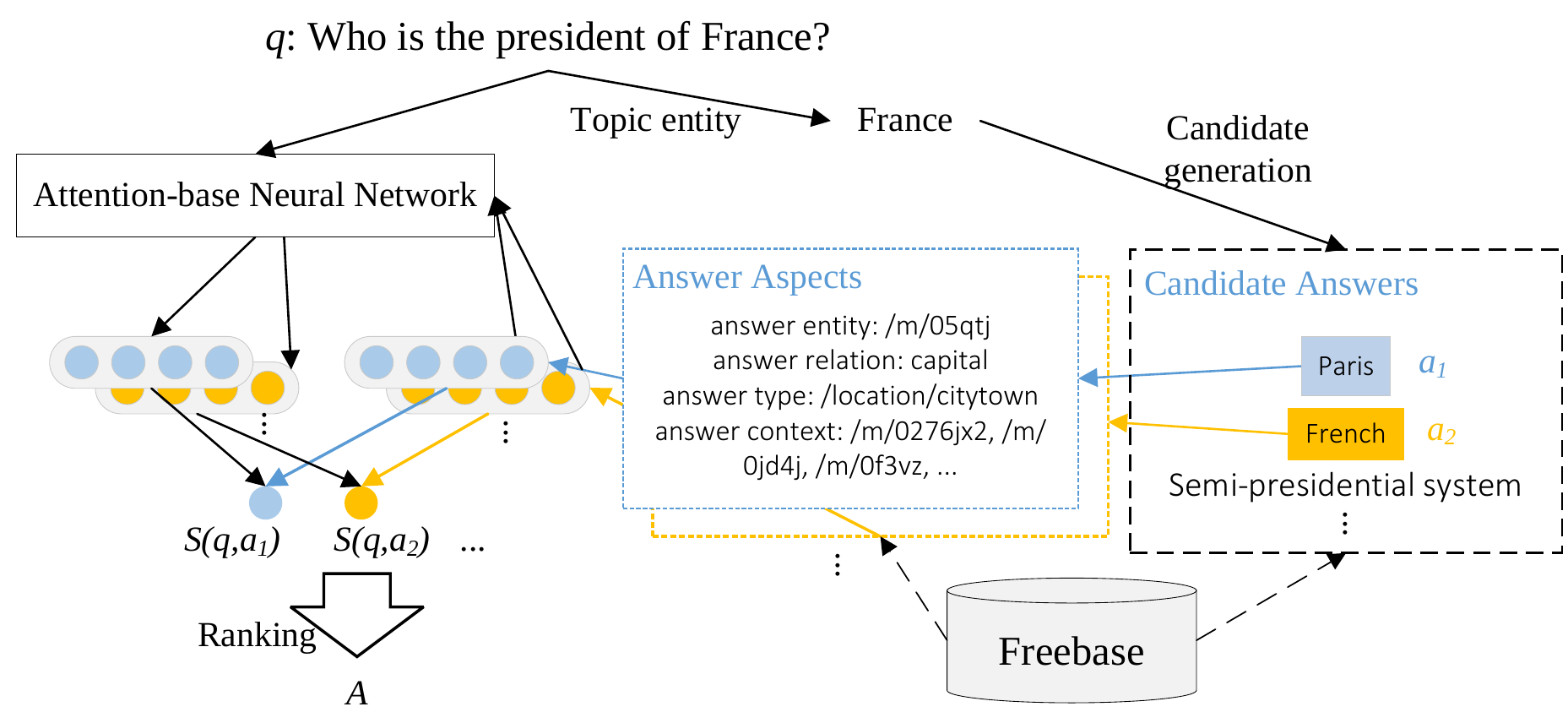}
	\caption{The overview of the proposed KB-QA system.}
	\label{fig:overview}
\end{figure}
The goal of the KB-QA task could be formulated as follows. Given a natural language question $q$, return an entity set $A$ as answers. The architecture of our proposed KB-QA system is shown in Figure \ref{fig:overview}, which illustrates the basic flow of our approach. First, we identify the topic entity of the question, and generate candidate answers from Freebase. Then, the candidate answers are represented with regard to their four aspects. Next, an attention-based neural network is employed to represent the question under the influence of the candidate answer aspects. Finally, the similarity score between the question and each corresponding candidate answer is calculated, and the candidates with the highest score will be selected as the final answers\footnote{We also adopt a margin strategy to obtain multiple answers for a question and this will be explained in the next section.}.

We utilize Freebase \cite{bollacker2008freebase} as our knowledge base. It now has more than 3 billion facts, and is used as the supporting KB for many QA tasks. In Freebase, the facts are represented by \textit{subject-property-object} triples {\tt (s,p,o)}. For clarity, we call each basic element a resource, which could be either an entity or a relation. For example, (\texttt{/m/0f8l9c, location.country.capital, /m/05qtj})\footnote{Note that the Freebase prefixes are omitted for neatness.} describe the fact that the capital of France is Paris, \texttt{/m/0f8l9c} and \texttt{/m/05qtj} are entities denoting \texttt{France} and \texttt{Paris} respectively, and \texttt{location.country.capital} is a relation.

\section{Our Approach}
\subsection{Candidate Generation}
The candidate answers should be all the entities of Freebase ideally, but in practice, this is time consuming and not really necessary. For each question $q$, we can use Freebase API \cite{bollacker2008freebase} to identify a \textit{topic entity}, which could be simply understood as the main entity of the question. For example, {\tt France} is the topic entity of question ``Who is the president of France?''. Freebase API method is able to resolve as many as 86\% questions if we use the top1 result \cite{yao2014information}. After getting the topic entity, we collect all the entities directly connected to it and the ones connected with 2-hop\footnote{For example, \texttt({/m/0f8l9c, governing\_officials, government\_position\_held.office\_holder, /m/02qg4z}) is a 2-hop connection.}. These entities constitute a candidate set ${C_q}$ .

\subsection{The Proposed Neural Attention Model}
\begin{figure}[!t]
	\centering
	\includegraphics[width=1\linewidth]{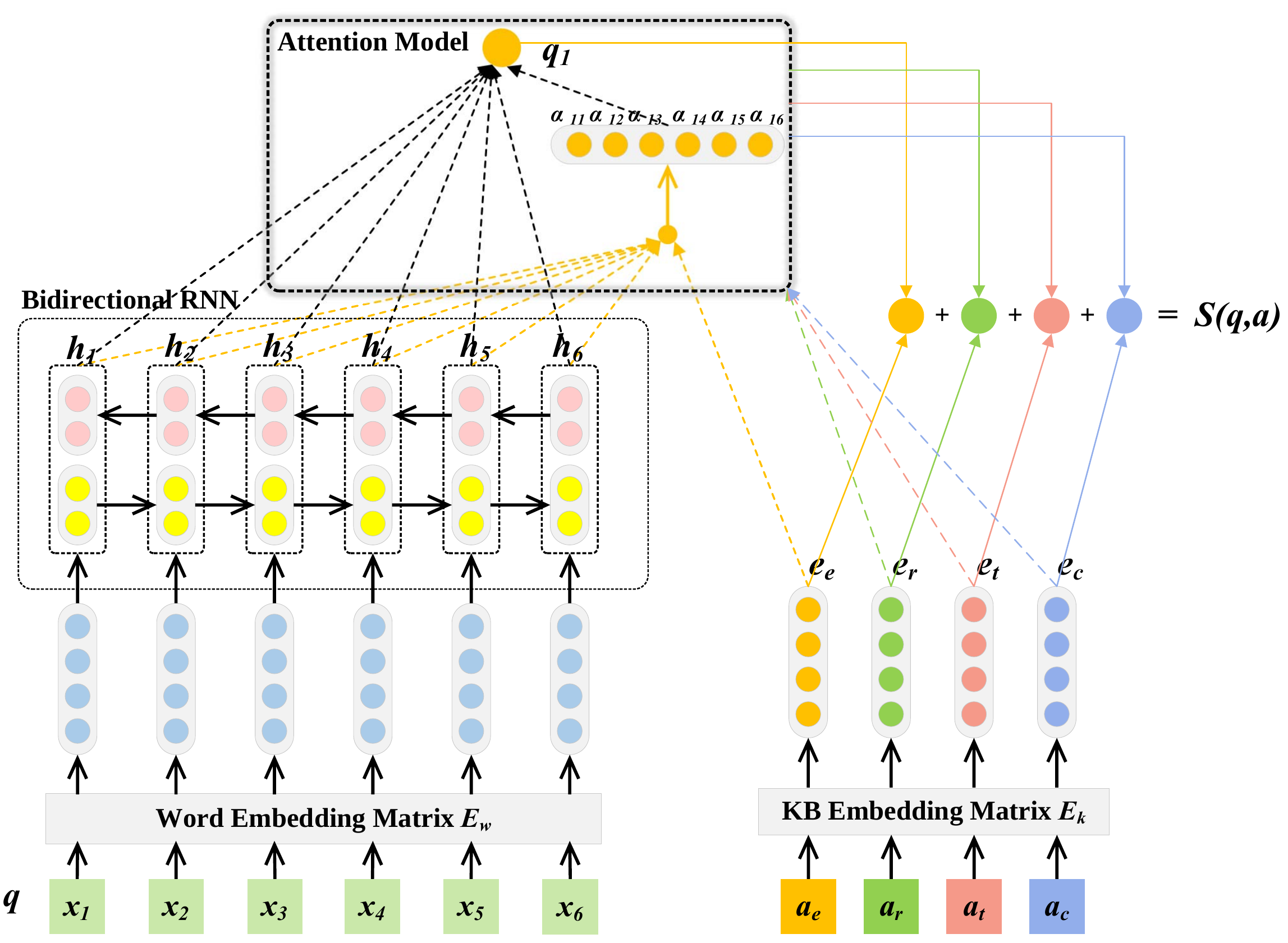}
	\caption{The architecture of the proposed attention-based neural network. Note that only one aspect (in orange color) is depicted for clarity. The other three aspects follow the same way.}
	\label{fig:main}
\end{figure}

We present an attention-based neural network, which represents the question dynamically according to different answer aspects. Concretely, each aspect of the answer pays different attention to the question and thus decides how the question is represented. The extent of the attention is used as the weight of each word in the question. Figure \ref{fig:main} is the architecture of our model. We will illustrate how the system works as follows.

\noindent\textbf{LSTM}

First of all, we have to obtain the representation of each word in the question. These representations retain all the information of the question, and could serve the following steps. Suppose question $q$ is expressed as $q=(x_1,x_2,...,x_n)$, where $x_i$ denotes the $i$th word. As shown in Figure \ref{fig:main}, we first look up a word embedding matrix ${E_w} \in \mathbb{R}{^{d \times {v_w}}}$ to get the word embeddings, which is randomly initialized, and updated during the training process. Here, $d$ means the dimension of the embeddings and $v_w$ denotes the vocabulary size of natural language words. Then, the embeddings are fed into a long short-term memory (LSTM) \cite{hochreiter1997long} networks. LSTM has been proven to be effective in many natural language processing (NLP) tasks such as machine translation \cite{sutskever2014sequence} and dependency parsing \cite{dyer2015transition}, and it is adept in harnessing long sentences. Note that if we use unidirectional LSTM, the outcome of a specific word contains only the information of the words before it, whereas the words after it is not taken into account. To avoid this, we employ bidirectional LSTM as \cite{bahdanau2015neural} does, which consists of both forward and backward networks. The forward LSTM handles the question from left to right, and the backward LSTM processes in the reverse order. Thus, we could acquire two hidden state sequences, one from the forward one $({\vec h_1},{\vec h_2},...,{\vec h_n})$ and the other from the backward one $({\mathord{\buildrel{\lower3pt\hbox{$\scriptscriptstyle\leftarrow$}}
\over h} _1},{\mathord{\buildrel{\lower3pt\hbox{$\scriptscriptstyle\leftarrow$}} \over h} _2},...,{\mathord{\buildrel{\lower3pt\hbox{$\scriptscriptstyle\leftarrow$}} \over h} _n})$. We concatenate the forward hidden state and the backward hidden state of each word, resulting in ${h_j} = [{\vec h_j};{\mathord{\buildrel{\lower3pt\hbox{$\scriptscriptstyle\leftarrow$}} \over h} _j}]$. The hidden unit of forward and backward LSTM is $\frac {d}{2}$, so the concatenated vector is of dimension $d$. In this way, we obtain the representation of each word in the question.

\noindent\textbf{Answer aspect representation}

In the answer end, we directly use the embedding for each answer aspect through the KB embedding matrix ${E_k} \in \mathbb{R}{^{d \times {v_k}}}$. Here, $v_k$ means the vocabulary size of the KB resources. The embedding matrix is randomly initialized and learned during training, and could be further enhanced with the help of the global information as described in Section 3.3. Concretely, we employ four kinds of answer aspects, namely, \textit{answer entity $a_e$}, \textit{answer relation $a_r$}, \textit{answer type $a_t$} and \textit{answer context\footnote{Here, the entities that directly connected to the answer entity is regarded as the answer context.} $a_c$}. Their embeddings are denoted as $e_e$, $e_r$, $e_t$ and $e_c$ respectively. It is worth noting that the answer context consists of multiple KB resources, and we denote it as $(c_1,c_2,...,c_n)$. We first acquire their KB embeddings $(ec_1,ec_2,...,ec_n)$ through $E_k$, then calculate an average embedding by $e_c = \frac {1}{n} \sum\limits_{i=1}^{n} {ec_i}$.

\noindent\textbf{Attention model}

The most crucial part of the proposed approach is the attention mechanism. Based on our assumption, each answer aspect should have different attention towards the same question. The extent of attention can be measured by the relatedness between each word representation $h_j$ and an answer aspect embedding $e_i$. We propose the following formulas to calculate the weights.
\begin{equation}
{\alpha _{ij}} = \frac{{\exp ({w_{ij}})}}{{\sum\limits_{k = 1}^{{n}} {\exp ({w_{ik}})} }}
\end{equation}
\begin{equation}
{w_{ij}} = {W^T}(tanh [h_j;e_i]) + b
\end{equation}
Here, ${\alpha _{ij}}$ denotes the attention weight of the $j$th word in the question, in terms of answer aspect $e_i$, where $e_i \in \{e_e, e_r, e_t, e_c\}$. $n$ is the length of the question. $W \in \mathbb{R}{^ {2d \times 1}}$ is an intermediate matrix and $b$ is an offset value. Both of them are randomly initialized and updated during training. Subsequently, the attention weights are employed to calculate a weighted sum of the words, resulting in a semantic vector that represent the question, according to the specific answer aspect $e_i$.
\begin{equation}
{q_i} = \sum\limits_{j = 1}^{{n}} {{\alpha _{ij}}{h_j}}
\end{equation}
By now, the similarity score of question $q$ and this particular candidate answer $a$ could be defined as follows.
\begin{equation}
S(q,a) = \sum\limits_{e_i \in \{e_e, e_r, e_t, e_c\}} {{q_i} \cdot {e_i}}
\end{equation}

The proposed attention model could also be intuitively interpreted as a re-reading mechanism \cite{hermann2015teaching}. Our aim is to select correct answers from a candidate set. When we consider a candidate answer, suppose we first look at its type, and we will re-read the question to find out which part of the question should be more focused (handling attention). Then we go to next aspect and re-read the question again, until the all the aspects are utilized. We believe that this mechanism is beneficial for the system to better understand the question with the help of the answer aspects, and leads to a performance promotion.

\noindent\textbf{Training}

We first construct the training data. Since we have question-answer pairs $(q,a)$ as supervision data, candidate set ${C_q}$ of question $q$ can be divided into two subsets, namely, correct answer set ${P_q}$ and wrong answer set ${N_q}$. For each correct answer $a \in {P_q}$, we randomly select $k$ wrong answers $a' \in {N_q}$ as negative examples. For some topic entities, there may be not enough wrong answers to acquire $k$ wrong answers. Under this circumstance, we extend $N_q$ from other randomly selected candidate set $C_{q'}$. With the generated training data, we are able to make use of pairwise training.

The training loss is given as follows.
\begin{equation}
L_{q,a,a'} = {[\gamma + S(q,a') - S(q,a)]_ + }
\end{equation}
Where $\gamma$ is a positive real number that ensure a margin between positive and negative examples. And ${[z]_ + }$ means $\max (0,z)$. The intuition of this training strategy is to guarantee the score of positive question-answer pairs be higher than negative ones with a margin.

The objective function is as follows.
\begin{equation}
\min \sum\limits_q {\frac{1}{{\left| {{P_q}} \right|}}} \sum\limits_{a \in {P_q}} {\sum\limits_{a' \in {N_q}} {{L_{q,a,a'}}} }
\end{equation}
We adopt stochastic gradient descent (SGD) to implement the learning process, mini-batches are utilized.

\noindent\textbf{Inference}

In testing stage, we straightforwardly take advantage of the candidate answer set $C_q$ of the question. We have to calculate $S(q,a)$  for each $a \in {C_q}$, and find out the maximum value $S_{max}$.
\begin{equation}
S_{max} = \mathop {\arg \max }\limits_{a \in {C_q}} \{ S(q,a)\}
\end{equation}
It is worth noting that many questions have more than one answer, so it is improper to set $S_{max}$ as the final answer. Instead, we make use of the margin $\gamma$ in the loss function, if the score of an candidate answer is within the margin compared with $s_{max}$, we put it in the final answer set.

\begin{equation}
A = \{ \hat a|{S_{max }} - S(q,\hat a) < \gamma\}
\end{equation}

\subsection{Combining Global Knowledge Information}
In this section, we elaborate how the global information of the KB could be leveraged. As stated before, we try to take into account the complete structural information of the KB. To this end, we adopt TransE model \cite{bordes2013translating} to represent the KB, and integrate the representations into the QA training process.


In TransE model, the entities and relations are represented by low dimensional embeddings. The basic idea is that the relations are regarded as translations in the embedding space. Here, for consistency, we denote each fact as $(s,p,o)$, and use boldface $(\bm{s,p,o})$  to denote their embeddings. The embedding of the tail entity $\bm{o}$ should be close to the embedding of head entity $\bm{s}$ plus the embedding of relation $\bm{p}$, i.e., $(\bm{s + p\approx o})$. The energy of a triple $(s,p,o)$ is equal to $d(s+p,o)$ for some dissimilarity $d$, defined as $\left\| \bm{{s + p - o}} \right\|_2^2$. To learn the embeddings, TransE minimizes the following loss function.

\begin{equation}
L_k = \sum\limits_{(s,p,o) \in S} {\sum\limits_{(s',p,o') \in S'} {{{[\gamma_k  + d(s + p,o) - d(s' + p,o')]}_ + }} }
\end{equation}

Where $S$ is the set of KB facts and $S'$ is the corrupted facts, which is composed of positive facts with either the head or tail replace by a random entity. The loss function favors lower values of the energy for positive facts than for negative facts.

In our implementation, we filter out the completely unrelated facts to save time. To be more specific, we first collect all the topic entities of all the questions as initial set. Then, we expand the set by adding direct connected and 2-hop entities. Finally, all the facts in which these entities appeared form the positive set. The negative facts are randomly corrupted ones. This a compromise solution due to the large scale of Freebase.

To combine the global information to our training process, we adopts a multi-task training strategy. Specifically, we perform our KB-QA training and TransE training in turn. After each epoch of KB-QA training, 100 epochs of TransE training is conducted, and the embeddings of the KB resources are shared and updated during both training processes. The proposed training process ensures that the global KB information act as additional supervision, and the interconnections among the resources are fully considered. In addition, as more KB resources are involved, the OOV problem will be relieved, which is able to bring additional benefits to the attention model.

\section{Experiments}
\subsection{Datasets}
To evaluate the proposed method, we select WEBQUESTIONS \cite{berant2013semantic} dataset that includes 3,778 question-answer pairs for training and 2,032 for testing. The questions are collected from Google Suggest API, and the answers are labeled manually by Amazon MTurk. All the answers are from Freebase. We use three-quarter (2,833) of the training data as training set, and the remaining quarter as validate set. F$_1$ score computed by the script provided by \cite{berant2013semantic} is select as the evaluation metric

\subsection{Settings}
For KB-QA training, we use mini-batch stochastic gradient descent to minimize the pairwise training loss. The mini-batch size is set to 50. The learning rate is set to 0.01. Both the word embedding matrix $E_w$ and KB embedding matrix $E_v$ are normalized after each epoch. The embedding size $d=128$, and the hidden unit size is 64. Margin $\gamma$ is set to 0.6. Negative example number $k=500$. The TransE training process defines the embeddings dimension to 128, and the mini-batch size is also 50. $\gamma_k$ is set to 1. All these hyperparameters of the proposed network is determined according to the performance on the validate set.

\subsection{Results}
\textbf{The Effectiveness of the proposed approach}

To demonstrate the effectiveness of the proposed approach, we compare our method with previous NN-based methods. Table \ref{tab:result1} shows the results on WEBQUESTIONS test set. The methods listed in the table all employ neural network for KB-QA. \cite{bordes2014open} applies BOW method to obtain a single vector for both questions and answers. \cite{bordes2014question} further improves their work by proposing the concept of subgraph embeddings. Besides the answer path, the subgraph contains all the entities and relations connected to the answer entity. The final vector is also obtained by BOW strategy. \cite{yang2014joint} follows the SP-based manner, but uses embeddings to map entities and relations into KB resources, then the question can be converted into logical forms. They jointly consider the two mapping process. \cite{dong2015question} uses three columns of CNNs to represent questions corresponding to three aspects of the answers, namely the answer context, the answer path and the answer type. \cite{bordes2015large} puts KB-QA into the memory networks \cite{sukhbaatar2015end} framework, and achieves the state-of-the-art performance. \textbf{ours} represents the proposed approach.

\begin{table}[!htb]
    \small
	\begin{center}
		\begin{tabular}{lc}
			\hline Method & F${_1}$ \\
            \hline
            \hline Bordes et al., 2014b & 29.7 \\
            Bordes et al., 2014a & 39.2 \\
            Yang et al., 2014 & 41.3 \\
            Dong et al., 2015 & 40.8 \\
            Bordes et al., 2015 & 42.2 \\
            \hline
            \textbf{ours} & \textbf{42.6} \\
            \hline
		\end{tabular}
		\caption{\label{tab:result1} The evaluation results on WEBQUESTIONS.}
	\end{center}	
\end{table}

From the results, we can observe that \textbf{ours} achieves the best performance on WEBQUESTIONS. Here \cite{bordes2014open,bordes2014question,bordes2015large} all utilize BOW model to represent the questions, while ours takes advantage of the attention of answer aspects to dynamically represent the questions. Also note that \cite{bordes2015large} uses additional training data such as Reverb \cite{fader2011identifying} and their original dataset SimpleQuestions. \cite{dong2015question} employs three fixed CNNs to represent questions, while ours is able to express the focus of each unique answer aspect in the question representation. Besides, the global KB information is leveraged. So, we believe that the results faithfully show that the proposed approach is more effective than the other competitive methods. It is worth noting that \cite{yih2015semantic} achieves an F${_1}$ of 52.5, much higher than other methods. Their staged system is able to address more questions with constraints and aggregations. However, their approach applies numbers of manually designed rules and features, which come from the observations on the training set questions. These particular manual efforts reduce the adaptability of their approach.

\noindent\textbf{Model Analysis}

In this part, we further discuss the impacts of the components of our model. Table \ref{tab:result2} indicates the effectiveness of different parts in the model.

\begin{table}[!htb]
    \small
	\begin{center}
		\begin{tabular}{lc}
			\hline Method & F${_1}$ \\
            \hline
            \hline
            LSTM & 38.2 \\
            Bi\_LSTM & 38.9 \\
            Bi\_LSTM + ATT & 41.6\\
            Bi\_LSTM + GKI & 40.4\\
            Bi\_LSTM + ATT + GKI & 42.6 \\
            \hline
		\end{tabular}
		\caption{\label{tab:result2} The results of different models.}
	\end{center}	
\end{table}

\textit{LSTM} employs unidirectional LSTM, and uses the last hidden state as the question representation. \textit{Bi\_LSTM} adopts a bidirectional LSTM. If we use $({\vec h_1},{\vec h_2},...,{\vec h_n})$ to denote the forward LSTM, and use $({\mathord{\buildrel{\lower3pt\hbox{$\scriptscriptstyle\leftarrow$}}
\over h} _1},{\mathord{\buildrel{\lower3pt\hbox{$\scriptscriptstyle\leftarrow$}} \over h} _2},...,{\mathord{\buildrel{\lower3pt\hbox{$\scriptscriptstyle\leftarrow$}} \over h} _n})$ to indicate backward LSTM, then the final presentation of the question is $[{\vec h_n};{\mathord{\buildrel{\lower3pt\hbox{$\scriptscriptstyle\leftarrow$}}\over h} _1}]$. \textit{Bi\_LSTM+ATT} is the bidirectional LSTM with neural attention (four answer aspects are used). \textit{Bi\_LSTM+GKI} denote the bidirectional LSTM model with global KB information (GKI). \textit{Bi\_LSTMS+ATT+GKI} is the same as \textbf{ours}, which is the bidirectional LSTM model with both attention model and global KB information.

From the results, we could observe the followings.

1) \textit{Bi\_LSTM+ATT} dramatically improves the F${_1}$ score by 2.7\% compared with \textit{Bi\_LSTM}. Similarly, \textit{ Bi\_LSTM+ATT+GKI} significantly outperforms \textit{ Bi\_LSTM+ GKI} by 2.2\%. They straightforwardly prove that the proposed attention model is effective.

2) \textit{Bi\_LSTM+GKI} performs better than \textit{Bi\_LSTM}, and achieves a 1.5\% improvement. Similarly, \textit{Bi\_LSTM+ATT+GKI} improves \textit{Bi\_LSTM+ATT} by 1\%. The results indicate that the proposed training strategy successfully leverages the global information of the underlying KB.

3) \textit{Bi\_LSTM+ATT+GKI} achieves the best performance as we expected, and  improves the original \textit{Bi\_LSTM} dramatically by 3.7\%. This directly shows the power of the attention model and the global KB information.

\begin{figure}[!h]
	\centering
	\includegraphics[width=0.98\linewidth]{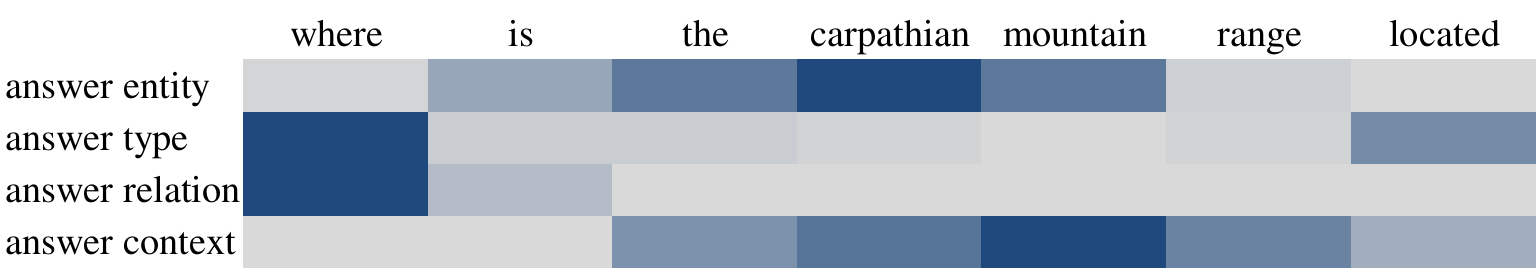}
	\caption{The visualized attention heat map. Answer entity: \texttt{/m/06npd (Slovakia)}, answer relation: \texttt{partially\_containedby}, answer type: \texttt{/location/country}, answer context: \texttt{(/m/04dq9kf, /m/01mp,...)}}
	\label{fig:heatmap}
\end{figure}

To clearly demonstrate the effectiveness of the attention mechanism in our approach, we present the attention weights of a question in the form of heat maps as shown in Figure \ref{fig:heatmap}.

From this example we can observe that our methods is able to capture the attention properly. It is instructive to figure out the attention part of the question when dealing with different answer aspects. The heat map will help us understand which parts are most useful for selecting correct answers. For instance, from Figure \ref{fig:heatmap}, we can see that \texttt{location.country} is paying great attention to ``Where'', indicating that ``Where'' is much more important than the other parts in the question when dealing with this type. In other words, the other parts are not that crucial since `Where'' is strongly implying that the question is asking about a location.

\subsection{Error Analysis}
We randomly sample 100 imperfectly answered questions and categorize the errors into two main classes as follows.

\noindent\textbf{Wrong attention}

In some occasions (17 in 100 questions, 17\%), we find the generated attention weights unreasonable. For instance, for question ``What are the songs that Justin Bieber wrote?'', answer type {\tt /music/composition} pays the most attention on ``What'' rather than ``songs''. We think this is due to the bias of the training data, and we believe these errors could be solved by introducing more instructive training data in the future.

\noindent\textbf{Complex questions and label errors}

Another challenging problem is the complex questions (34\%). For example, ``When was the last time Knicks won the championship?'' is actually to ask the last championship, but the predicted answers give all the championships. This is due to that the model cannot learn what does ``last'' mean in the training process. In addition, the label mistakes also influence the evaluation (3\%) . For example, ``What college did John Nash teach at?''. The labeled answer is {\tt Princeton University}, but {\tt Massachusetts Institute of Technology} should also be an answer, and the proposed method is able to answer it correctly.

Other errors include topic entity generation error and the multiple answers error (giving more answers than expected). We guess these errors are caused by the simplest implementations of the related steps in our method, and we will not explain them in detail due to space limitation.

\section{Related Work}
\subsection{Neural Network-based KB-QA}
\cite{bordes2014open} first applies NN-based method to solve KB-QA problem. The questions and KB triples are represented by vectors in a low dimensional space. Thus the cosine similarity could be used to find the most possible answer. BOW method is employed to obtain a single vector for both the questions and the answers. Pairwise training is utilized, and the negative examples are randomly selected from the KB facts. They also present a training data generation method, i.e., using KB facts to and some heuristics rules to generate natural language questions.

\cite{bordes2014question} further improves their work by proposing the concept of subgraph embeddings. The key idea is to involve as much as information in the answer end. Besides the answer triple, the subgraph contains all the entities and relations connected to the answer entity. The final vector is also obtained by BOW strategy.

\cite{yih2014semantic} focuses on single-relation questions. The KB-QA task is divided into two parts, i.e., finding the entity mention-entity mapping and then mapping the remaining relation pattern to the KB relation. They train two CNN models to perform the mapping processes.
\cite{yang2014joint} handles entity and relation mapping as joint procedures. Strictly speaking, these two methods follow the SP-based manner, but they take advantage of neural networks to obtain intermediate mapping results.

The most similar work to ours is \cite{dong2015question}. They consider the different aspects of answers, using three columns of CNNs to represent questions respectively. The difference is that our approach uses attention mechanism for each unique answer aspect, so the question representation is not fixed to only three types. Moreover, we utilize the global KB information.

\subsection{Attention-based Model}
\cite{bahdanau2015neural} first applies attention model in NLP. They improve the encoder-decoder Neural Machine Translation (NMT) framework by jointly learning alignment and translation. They argue that representing the source sentence by a fixed vector is unreasonable, and propose a soft-align method, which could be understood as the attention mechanism.

\cite{luong2015effective} is also tackling machine translation task. They propose two attentions models, i.e., a global model and a local model. The latter further indicates a small scope to attend, and achieves better results.

\cite{rush2015neural} implements sentence-level summarization task. They utilize local attention-based model that generate each word of the summary conditioned on the input sentence.

Our approach differs from previous work in that we are using attentions to help represent question dynamically, not generating current word from vocabulary as before.

\section{Conclusion}
In this paper, we focus on the KB-QA task. First, we consider the impacts of the different answers and their aspects when representing the question, and propose a novel attention-based model for KB-QA. Specifically, the attention of the answer aspect for each word in the question is used. This kind of dynamic representation is more precise and flexible. Second, we leverage the global KB information, which could take full advantage of the complete KB, and also could alleviate the OOV problem. The extensive experiments demonstrate that the proposed approach could achieve better performance compared with other state-of-the-art NN-based methods.

\bibliographystyle{named}
\bibliography{arXiv}

\end{document}